\documentclass{iaus}
\usepackage{graphicx}

\title[Spheroids scaling relations] 
{ Spheroids scaling relations over cosmic time}

\author[T.Treu]   
{Tommaso Treu$^1$%
\affiliation{$^1$Department of Physics, University of California, Santa Barbara, CA 93106-9530, USA \break email:tt@physics.ucsb.edu}}

\pubyear{2006}
\volume{235}  
\pagerange{119--126}
\date{?? and in revised form ??}
\jname{Galaxy Evolution across the Hubble Time}
\editors{F.Combes \& J. Palous, eds.}
\newcommand{\apj}{{\it ApJ}}
\newcommand{\mnras}{{\it MNRAS}}
\newcommand{\kms}{kms$^{-1}$}
\begin{document}

\maketitle

\begin{abstract}
I report on recent measurements of two scaling relations of spheroids
in the distant universe: the Fundamental Plane, and the relation
between lensing velocity dispersion and stellar velocity
dispersion. The joint analysis of the two scaling relations indicates
that the most massive (above $\sim10^{11.5} M_{\odot}$) spheroids are
consistent with no evolution since $z\sim1$ both in terms of star
formation and internal structure. Furthermore their total mass density
profile is on average well described by an isothermal sphere with no
evidence for redshift evolution. At smaller masses the picture appears
to be substantially different, as indicated by evidence for
substantial recent star formation (as much as 20-40\% of stellar mass
formed since $z\sim1$), and by hints of a reduced dark matter content
at smaller masses. A larger sample of lenses extending to velocity
dispersions below 200\kms, and to redshifts above $>0.5$ is needed to
verify these trends.

\keywords{gravitational lensing; stellar dynamics; galaxies:
elliptical and lenticular, cD; galaxies: evolution galaxies:
formation; galaxies: halos; (cosmology:) dark matter}
\end{abstract}

\firstsection 
\section{Introduction}

Spheroids (i.e. elliptical and lenticular galaxies, or collectively
early-type galaxies) are observed to obey tight empirical scaling
relations, i.e. correlations between observable properties. Well known
examples of tight scaling relations are the Fundamental Plane (Dressler et
al. 1987; Djorgovski \& Davis 1987; hereafter FP) and the M$_{\rm
BH}$-$\sigma$ relation (Ferrarese \& Merritt 2001; Gebhardt et al.\
2001). The connection between observables involving dynamics, stellar
populations, and the mass of the central black hole, indicates that
mass assembly, star formation history, and nuclear activity are
interconnected.

A common application of scaling relations in the distant universe is their
use as ``generalized standard rods''. For example the FP is often used
to derive the evolution of the effective mass to light ratio as a
function of mass and redshift (Franx 1993; Treu et al.\
1999). Assuming pure luminosity evolution this can then be converted
into a star formation history (Treu et al. 2005a,b). Similarly the
evolution of the M$_{\rm BH}$-$\sigma$ relation can be used to
quantify the relative growth of bulges and black holes over cosmic
time (e.g. Shields et al.\ 2003; Woo et al.\ 2006).  Here I will
briefly report on recent progress in two areas: the evolution of the
Fundamental Plane, and that of a less known scaling relation, i.e. the
correlation between stellar velocity dispersion and total mass as
measured by strong gravitational lensing. 

\section{The Fundamental Plane}

By defining an effective mass $M$ in the terms of stellar velocity
dispersion and effective radius ($M={\rm K_V} \sigma^2R_{\rm e}$,
where ${\rm K_V}$ is the so-called virial coefficient) the FP can be
seen as a scaling relation between $M$ and the effective mass-to-light
ratio $M/L$. If one assumes that $M/L$ is proportional to the stellar
mass-to-light ratio $M*/L$, the FP provides a very robust measurement
of the evolution of $M*/L$ as a function of $M$. Note that this
procedure does not assume a constant virial coefficient as a function
of mass, but only that the virial coefficient be constant as a
function of time at any given mass.

\begin{figure}
\includegraphics[height=2.7in,angle=0]{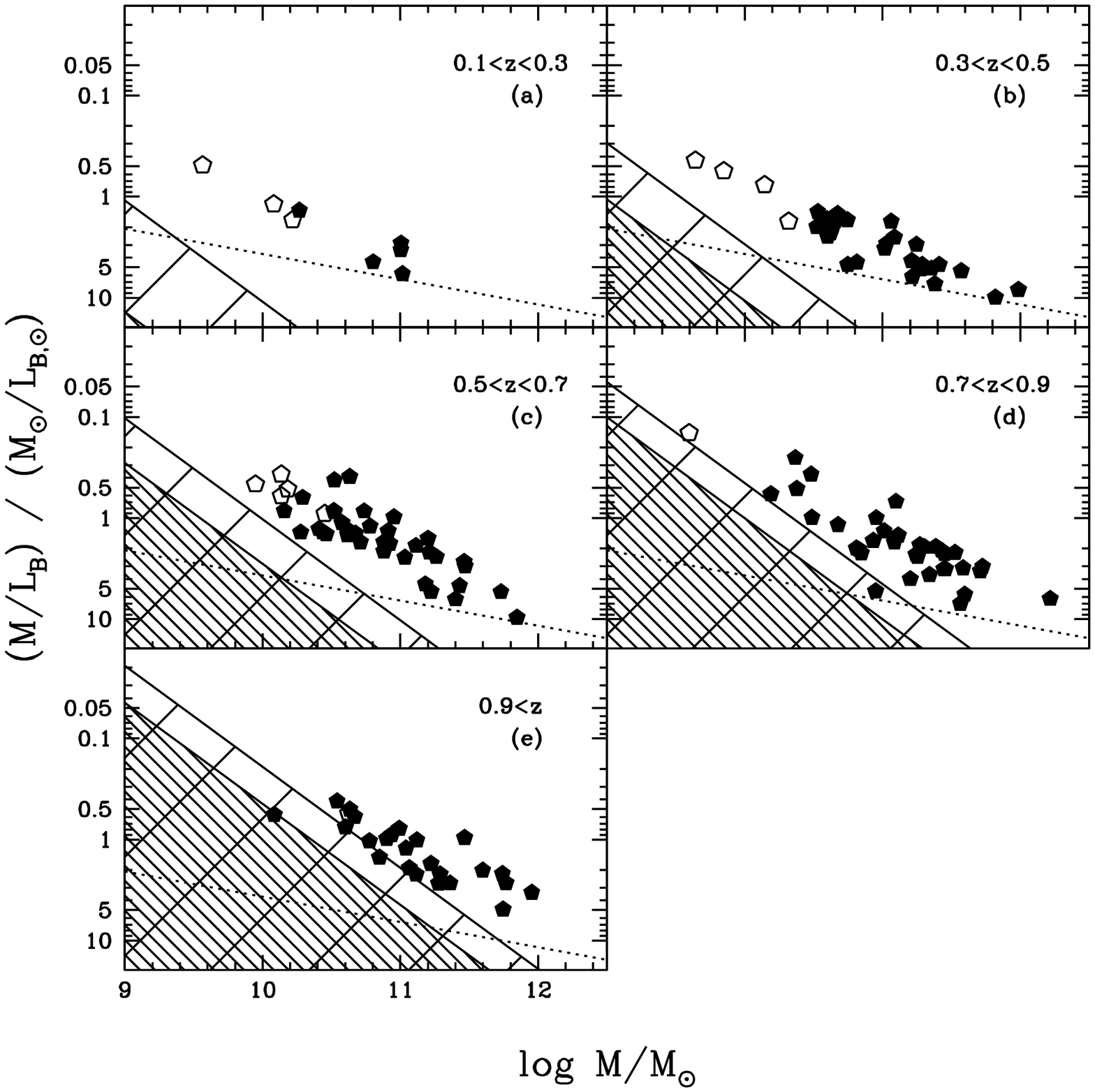}
\includegraphics[height=2.7in,angle=0]{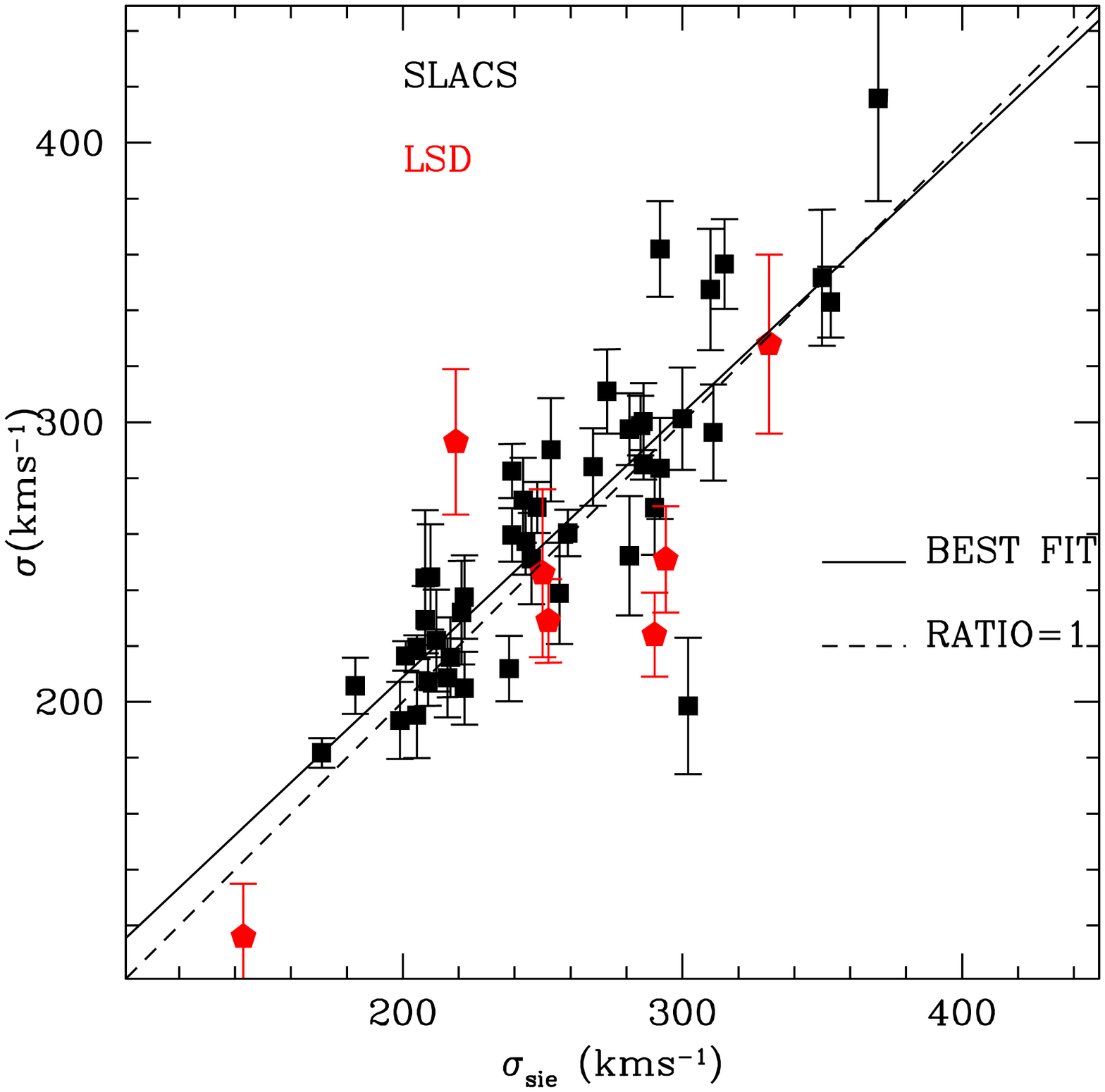}
\caption{Left: Fundamental Plane as a function of redshift projected
on the $M$-$M/L$ plane. The local relation is shown as a dotted line
for comparison. Hatched regions are excluded by the magnitude
limit. Although selection effects enhance the observed evolution at
the low mass end, careful modeling shows that the evolutionary rate
depends on mass, and that the intrinsic scatter of the FP increases
with redshift. From Treu et al. 2005b.
Right: Scaling relation between velocity dispersion of the stars
($\sigma$), and that of the singular velocity dispersion ellipsoids
that best fits the lensing constraints ($\sigma_{\rm SIE}$), for the
LSD and SLACS sample as of August 2006. \label{fig1}}
\end{figure}

Observational studies are now capable of probing out to redshift $\sim
1$ with samples of over 100 objects, covering more than an order of
magnitude in $M$. The main results are shown in Figure~\ref{fig1} (see
also van der Wel et al.\ 2005 and di Serego et al.\ 2005). The FP
evolves with redshift, in the sense that galaxies were brighter at
high-z than today for a given mass. Evolution is slowest for the most
massive objects, consistent with pure luminosity evolution if the most
massive galaxies have the oldest stellar populations. In terms of
assembly history, the fraction of stellar mass assembled below
$z\sim1$ is constrained to be of order 1\% for the most massive
objects ($M> 10^{11.5} M_{\odot}$), and up to $\sim20-40\%$ for
galaxies of masses of order $10^{10}$ M$_{\odot}$. This is another
manifestation of the concept of 'downsizing', sometimes called
'anti-hierarchical' behavior. Recent theoretical work suggests that
AGN feedback could reconcile this trend with the hierarchical
paradigm. Direct empirical tests of this mechanism are needed.

\section{A scaling relation measuring density profiles}

A recent explosion in the number of gravitational lenses with measured
stellar velocity dispersion $\sigma$ ($>50$; Bolton et al.\ 2005, 2006
and in prep.) allows one to study systematically the scaling of
$\sigma$ with $\sigma_{\rm SIE}$, i.e. the velocity dispersion of the
singular isothermal ellipsoid that best fits the lensing geometry. The
latter is a very robust, and almost model independent, measure of mass
within the critical line, i.e. approximately within a cylinder of
radius the Einstein Radius. This is typically in the range 0.5-5
$R_{\rm e}$. Since $\sigma$ measures the mass inside $\sim R_{\rm
e}/8$, this scaling relation effectively provides a measurement of the
slope of the mass density profile. Modulo small corrections due to
anisotropy, projection effects, and the precise location of the
Einstein Radius with respect to the effective radius, a ratio of unity
indicates an isothermal total mass density profile (see Treu \&
Koopmans 2002, 2004, Koopmans \& Treu 2003, Koopmans et al.\ 2006, for
detailed modeling). This is observed to be generally the case within
the errors, with few outliers (Figure~\ref{fig1}).

This tight scaling relation can be seen as the result of a
``bulge-halo conspiracy'', because neither of the two components has
an isothermal profile and yet the sum of the two is very close to
it. Remarkably, no significant evolution of the total mass density
slope is detected. Whatever mechanism produces the conspiracy it seems
to be at work at high-z for these massive galaxies. Note that SLACS
lens galaxies are indistinguishable from a control sample of
identically selected non-lens early-type galaxies within the current
observational uncertainties (Treu et al.\ 2006), and therefore these
results can be generalized to the entire population of massive
early-type galaxies. From an observational point of view, there is
room for improvement, especially at $z>0.5$, where the samples are
still pitifully small and SDSS runs out of lenses. A comparable sample
of $\sim 50$ lenses at $z>0.5$ would help enormously to understand
evolutionary trends.

%

\subsection{Does the dark matter content depend on mass?}

Finally, I will briefly discuss the fraction of dark matter as a
function of mass. There is convincing evidence that the most massive
early-type galaxies have dark matter halos more extended than the
stellar component, but the evidence is not quite as compelling for
less massive galaxies. Determining the relative contribution of dark
and stellar matter as a function of mass is important for a number of
reasons, e.g. understanding the efficiency of star formation as a
function of mass, and the interplay between baryons and dark matter.
In terms of scaling relations, one of the possible explanations of the
so-called ``tilt'' of the Fundamental Plane is a systematic increase
with effective mass of the dark matter fraction inside some fiducial
radius (e.g. Ciotti et al.\ 1996).

\begin{figure}
\includegraphics[height=2.7in,angle=0]{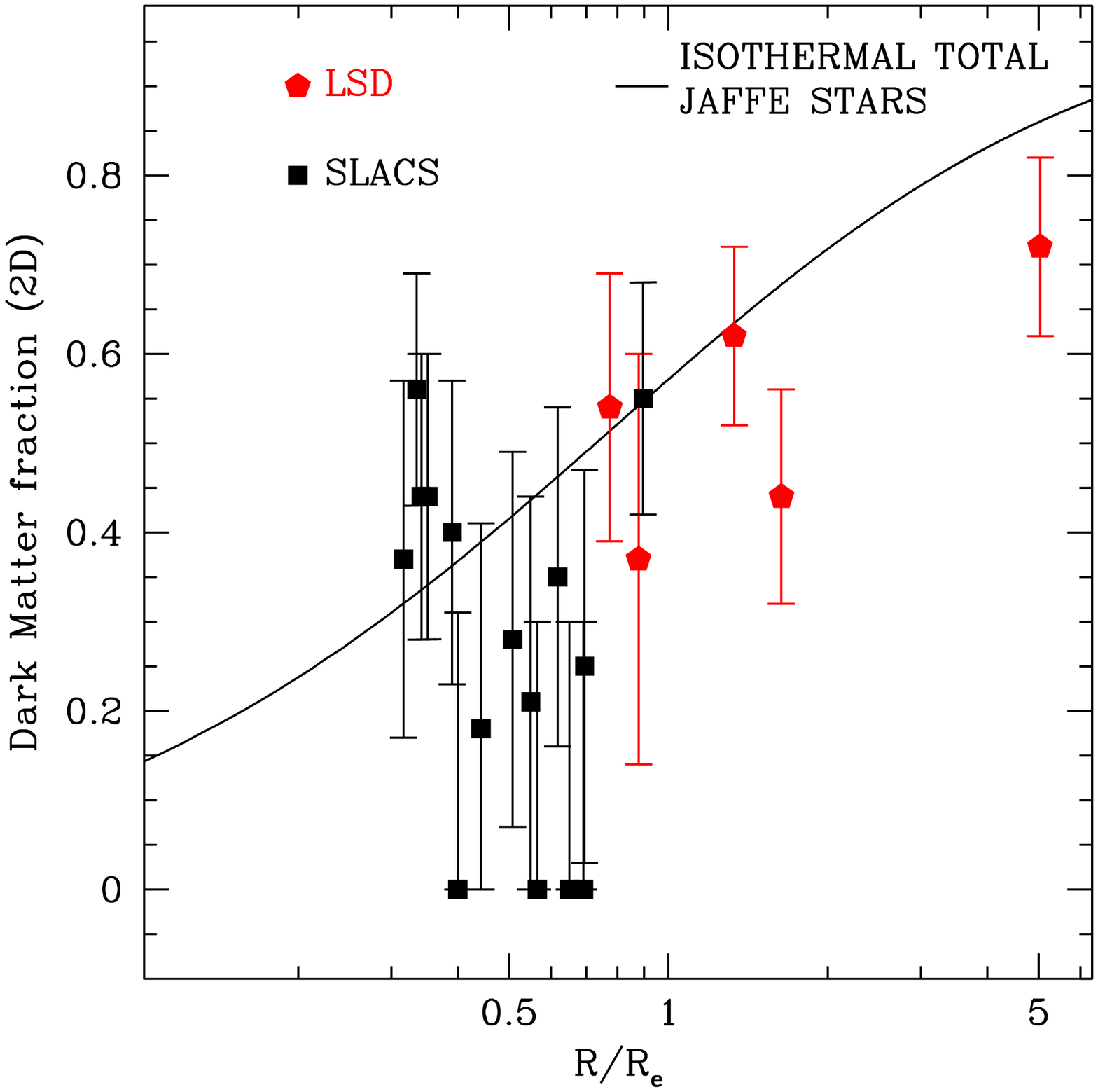}
\includegraphics[height=2.7in,angle=0]{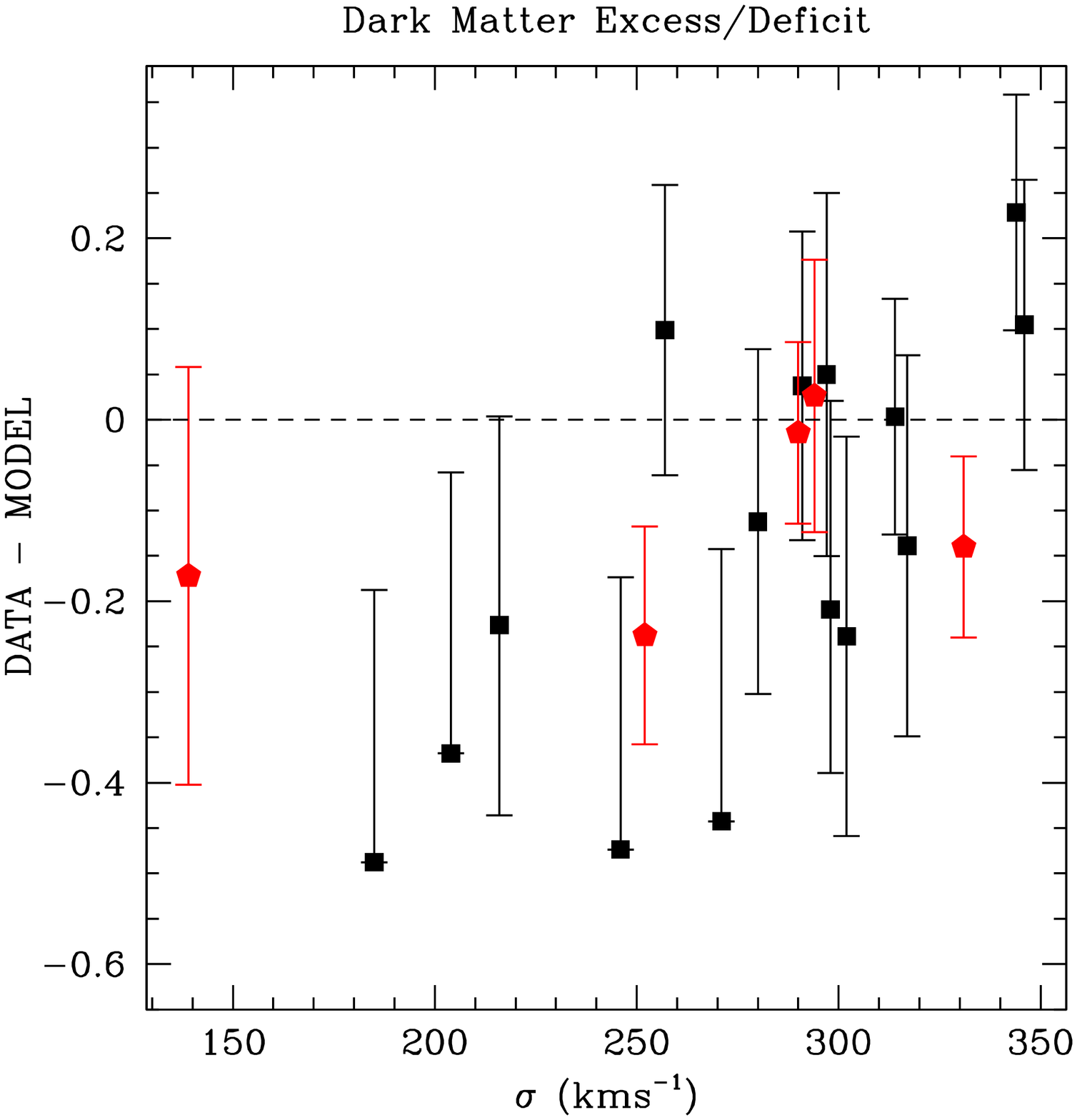}
  \caption{Left: Dark matter fraction inside the cylinder of radius
equal to the Einstein Radius compared to the expectation of a model
where the total mass distribution is isothermal and the stars follow a
Jaffe profile. Right: dark matter excess/deficit with respect to the
model. \label{fig:mass}}
\end{figure}

A joint lensing and dynamical analysis of the SLACS+LSD sample (Treu
\& Koopmans 2004; Koopmans et al. 2006) provides a robust measurement
of the fraction of dark matter at the Einstein Radius for each
lens. Since lenses span a range in velocity dispersions (a proxy for
mass) and in Einstein Radii (expressed in units of the effective
radius) this appears to be a promising way to attack this problem.
The promise of this method is illustrated in Figure~\ref{fig:mass}. In
the left panel, I show the fraction of dark matter inside the Einstein
Radius as derived from the lensing and dynamical analysis as a
function of the Einstein Radius in units of the effective radius. If
the dark matter halo is more spatially extended than the light, the
fraction of dark matter will be increasing with radius. Therefore,
this effect has to be taken into account, in order to study trends
with velocity dispersion.  One way to decouple the dependency on
radius from that on mass is to look at deviations from a simple
reference model, known to work quite well for the most massive systems
(e.g. Treu \& Koopmans 2002). This simple model consists of an
isothermal total mass profile, including a Jaffe (1983) profile
representing the stars (the remaining mass is assumed to be dark
matter). The Jaffe profile is normalized so that the fraction of dark
matter is zero at the center. Thus, the fraction of dark matter is a
universal function of radius in units of the effective radius, shown
by a solid line in the figure. In the right panel, I show the dark
matter deficit/excess with respect to the simple reference
model. Tantalizingly, it seems that the simple model works very well
for the most massive systems, while the least massive systems appear
to show a deficit of dark matter with respect to the model. Drawing
quantitative conclusions from this preliminary analysis is premature,
as a comprehensive analysis of a much enlarged sample of lenses is
underway.

%

\begin{acknowledgments}

I would like to acknowledge the many contributions to this work from
my friends and collaborators Richard Ellis, L\'eon Koopmans, Adam
Bolton, Scott Burles, and Leonidas Moustakas. This research is
supported by NASA through STScI grants GO-10174, GO-10494, AR-9920,
AR-09960. Many thanks to the AAS and the IAU for travel grants and to
the organizers of IAU235 for inviting me.

\end{acknowledgments}



\end{document}